# DIRECT DETERMINATION OF THE FULLY DIFFERENTIAL CROSS SECTION BY THE TIME-DEPENDENT WAVE FUNCTION

Zorigt Gombosuren[a], Khenmedekh Lochin[a], Altangoo Ochirbat[b] and Aldarmaa Chuluunbaatar[a,*]

[a]Department of Physics, Mongolian University of Science and Technology, Ulaanbaatar, Mongolia
[b]Department of Physics, Mongolian National University of Education, Ulaanbaatar, Mongolia

This paper focused on showing that the fully differential cross section of ionization during a collision of a proton and an antiproton with a hydrogen atom is directly expressed by the time-dependent wave function. For the projectile, wave function corresponding to the specific scattering angle was converted by two-dimensional Fourier transform from the wave functions of corresponding to impact parameters. This wave function shows how the ejected electron probability density distribution varies with time. We are shown that the calculation of the fully differential cross section of ionization can be directly determined by the local value of the wave function without the need to calculate the spatial integral for calculating the transition amplitude. It has been shown that the direct determination of the fully differential cross section by this time-dependent wave function is in good agreement with the results of determined the traditional method is by the transition amplitude.

## I. INTRODUCTION

One of the possible quantities measured in experiments involving charged particles collisions with atoms is the full differential cross section (FDCS). The works of Jones and Madison [1] and Voitkiw and Ullrich [2] show FDCS calculations for proton-hydrogen and antiproton-hydrogen collisions by perturbation theory. Recent years non-perturbation theoretical calculations have been developed by many researchers [3-9]. We numerically calculated the time-dependent Schrödinger equation (TDSE) to determine the electron wave function of the hydrogen atom. In this calculation, the projectile interaction is calculated using the impact parameter method, and the electron wave function is calculated using the discrete-variable representation of the Coulomb wave function (CWDVR) [ 11 - 13 ]. The time dependence of the electron probability density distribution corresponding to a certain impact parameter is presented in the papers [16 - 19]. Since these are not experimentally observable, we aim to make efficient calculations that are experimentally observable.

aldarmaa@must.edu.mn

We constructed a wave function corresponding to a specific ion scattering angle from the wave functions corresponding to the impact parameters by Fourier transform. Therefore, this wave function represents the electron probability density distribution corresponding to a particular projectile scattering angle. Moreover, the calculation results show that the FDCS can be directly determined by this wave function. Atomic units $\hbar = e = m_e = 1$ are used throughout the paper unless otherwise stated.

## II. THEORY

In the Schrödinger equation, let's calculate the interaction between a charged particle (projectile) and a hydrogen atom as a time-dependent potential. TDSE is written as follows.

$$i\frac{\partial}{\partial t}\Psi(\vec{r},t) = [\hat{H}_0 + \hat{V}(\vec{r},t)]\Psi(\vec{r},t) \quad (1)$$

$$\hat{V}(\vec{r},t) = \frac{-Z_p}{|\vec{R}(b,0,vt)-\vec{r}|}. \quad (2)$$

Here, $t$-is the time, $b$-is impact parameter, $v$-is the velocity of the projectile, $Z_p$ - projectile charge $\vec{R} = \vec{v}t + \vec{b}$. Figure 1 shows the kinematic scheme and parameters of the collision. Denotes $\Psi_{\vec{b}}(t,\vec{r})$ the wave function denotes the correspond to $b$-impact parameter.

Using the wave function calculated by the impact parameter method, the wave function corresponding to a certain scattering angle for the projectile is obtained by two-dimensional Fourier transform.

$$\Psi_{\vec{\eta}}(t,\vec{r}) = \frac{1}{2\pi}\int d\vec{b}\, e^{i\vec{\eta}\vec{b}} e^{i\delta(b)} \Psi_{\vec{b}}(t,\vec{r}) \quad (3)$$

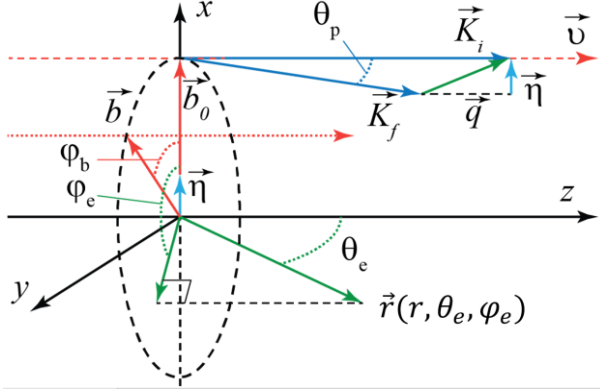

**Figure. 1**. Collision scheme of projectile ion and hydrogen atom. Projectile moves in a straight line along the z axis. $\vec{K}_i, \vec{K}_f$ – are the initial and final momentum of the projectile ion, $\vec{r}_e$- is the electron coordinate, $\vec{\eta}$ –is (perpendicular to $\vec{v}$) component of the projectile momentum transfer $\vec{q}$.

Here, $\delta(b)$ is the phase at which the nucleus of the hydrogen atoms interaction [10].

$$\delta(b) = \frac{2Z_a Z_p}{v} \cdot \ln(v \cdot b). \quad (4)$$

Atomic nucleus and projectile charge are correspond to $Z_a = 1$ ба $Z_p = \pm 1$ respectively. In equation (3), the wave function corresponding to $\vec{\eta}$. The $\vec{\eta}$ corrosponds to one scattering angle. Let us investigate the possibility of directly determining the FDCS by the wave function $\Psi_{\vec{\eta}}(t, \vec{r})$.

The electron probability density is written as follows:

$$\frac{d^3 P(\vec{\eta})}{dV d\vec{\eta}} = \left|\Psi_{\vec{\eta}}(t, \vec{r})\right|^2 \quad (5)$$

In this case, we write the volume differential $V$ in the spherical coordinate system. If ε is a function that depends on r, we can write the differential probability by shifting the variable as follows.

$$\frac{d^3 P(\vec{\eta})}{d\varepsilon d\Omega_e d\vec{\eta}} = \frac{\left|\Psi_{\vec{\eta}}(t,\vec{r})\right|^2 r^2 dr}{d\varepsilon}. \quad (6)$$

Here, $\varepsilon$-is ejection energy. In the relativistic coordinate system [5], when the scattering angle is small, FDCS is written as follows:

$$\frac{d^3 \sigma}{d\varepsilon d\Omega_e d\Omega_P} = K_i K_f \frac{\left|\Psi_{\vec{\eta}}(t,\vec{r})\right|^2 r^2}{d\varepsilon/dr}. \quad (7)$$

To calculate the FDCS with the above expression, it is necessary to determine the relationship between the ejection energy ε and the coordinates.

To solve this problem, the wave function can be written in the following form:

$$\left|\Psi_{\vec{\eta}}(t,\vec{r})\right| = a, \quad \arg\left(\Psi_{\vec{\eta}}(t,\vec{r})\right) = S \quad (8)$$

Here, the phase is $S$ and the modulus is $a$. By substituting it into the Schrödinger equation, the following quantum Hamilton-Jacob equation can be derived [15].

$$-\frac{\partial}{\partial t}S(\vec{r},t) = \frac{1}{2}\left(\nabla S(\vec{r},t)\right)^2 + U - \frac{1}{2a}\Delta a \quad (9)$$

Here $\nabla S(\vec{r}, t)$ on the right side of the equation is a particle momentum, $\frac{1}{2}\left(\nabla S(\vec{r}, t)\right)^2$ is the kinetic energy, the potential energy of the field $U$ and $-\frac{1}{2a}\Delta a$ are Bohm potential energy. The right side of equation 9 is the total energy of the electron. Therefore, the total energy of the electron is expressed as:.

$$\varepsilon(\vec{r}, t) = -\frac{\partial}{\partial t}S(\vec{r}, t) \quad (10)$$

The relationship between energy and coordinates and the radial derivative of energy can be determined by expression (10). (See Figure 5). Therefore, using the relation between coordinates and energy, it is possible to determine the FDCS by expression (7). traditionally FDCS in quantum mechanics expressed by the transition amplitude as follows [10].

$$\frac{d^3\sigma}{d\varepsilon d\Omega_e d\Omega_P} = K_i K_f \left|T(\varepsilon, \theta_e, \varphi_e, \eta, \varphi_\eta)\right|^2 \quad (11)$$

Here, ionization amplitude is:

$$T(\varepsilon, \theta_e, \varphi_e, \eta, \varphi_\eta) = \langle \Psi_{\vec{k}}^{(-)} | \Psi_{\vec{\eta}}(t, \vec{r}) \rangle. \quad (12)$$

This amplitude expression (12) is similar to the results described by other researchers [10]. (12) is confirmed to be the same as that of other researchers when the order of the electron coordinate integral and the integral of the impact parameter are changed.

## III. CALCULATION RESULTS

We will now show in a numerical example that the FDCS determined by the wave function (7) coincides with the FDCS calculated by the traditional method (11).

The TDSE was calculated by the CWDVR method when the antiproton or proton energies were 200 keV. In this method, the electron wave function is decomposed into a spherical harmonic function in a spherical coordinate system, and the radial function is calculated as a time-dependent function. The radial coordinates were discretized by the roots of the Coulomb wave function, and the pseudo-spectral basis functions of hydrogen atoms were obtained in accordance with it. Calculations were performed by decomposing the electron wave function with this pseudo-spectral basis functions. The notation of the parameters is taken as in [13]. For the antiproton, the charge number of the Coulomb wave function is Z = 120, the wave number is k = 8, the number of radial nodes is N = 1200, and rmax = 457.8.. The maximum value of the orbital quantum number is taken as 8 and the impact parameter $b_{max}$ = 30.8. The projectile z coordinate is from -80 to 1000 in steps of Δz = 0.08. For proton, the values of the parameters are Z = 120, k = 4, N = 800, $r_{max}$ = 583.1, $b_{max}$ = 37.5, $z_{max}$ = 560, and Δz = 0.16 respectively. In theoretical works [16-19] on the process of collision of protons and antiprotons with hydrogen atoms, the evolution of the wave function corresponding to the value of a certain impact parameter is described by the square of its modulus.

Although this represents the evolution of the collision process, it cannot be measured by experiment. This is because it is not yet possible to conduct experiments and measurements at the particular impact parameter . I n order to show the experimentally measured results, we derived the wave function corresponding to a specific scattering angle of the projectile by formula (3). This wave function allows us to see the evolution of the collision while representing quantities that can be measured experimentally. Figures 2 and 3 plot the square of the wave function module ($|\Psi_{\vec{\eta}}(t,\vec{r})|^2 r^2$) in the scattering plane. For this figures, the scattering angle of proton and antiproton is 0.0002 rad or $\eta = 0.519$. When the z-coordinates of the antiproton are 160, 240, 320 and the proton is 240, 320, 400, the times in the picture match. Figure 2 shows the probability distribution of ionized electrons moving outward.

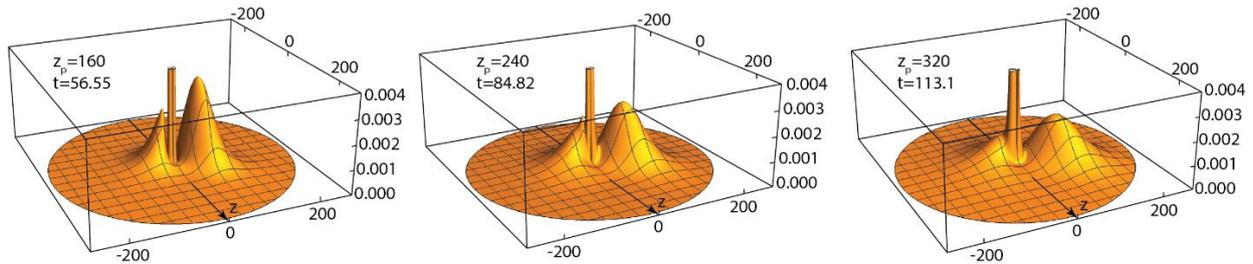

Figure 2. z coordinate $Z_p$ of antiproton and time t, electron probability density on the scattering plane

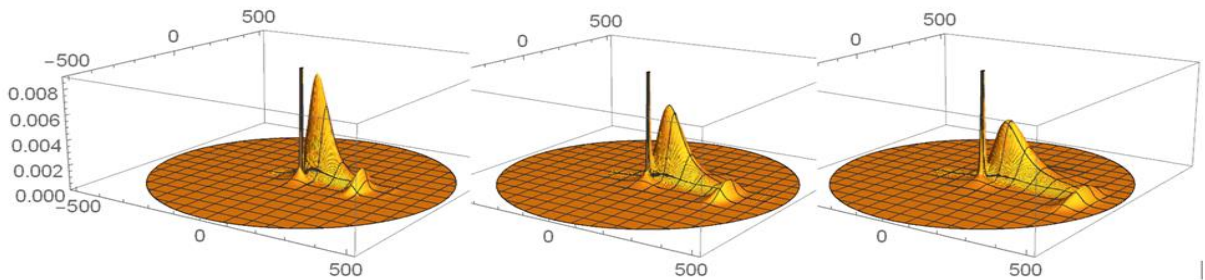

**Figure 3.** The probability density of electrons in the corresponding scattering plane when the z-coordinate of the proton is 240, 320, 400.

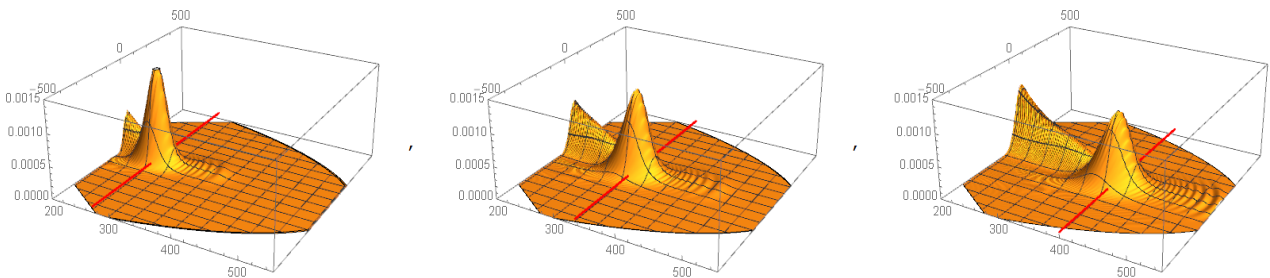

*Figure 4. The probability density of an electron in the plane of scattering corresponding to the moment of time when the z-coordinate of the proton is 272, 336, 400 (represented by red line).*

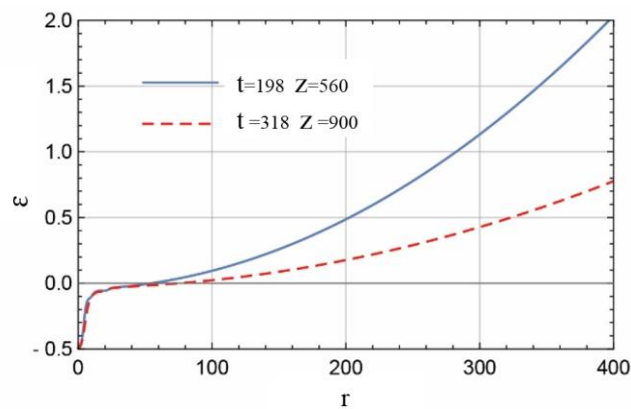

**Figure. 5** Coordinate and energy relation at 2 instants of time. Direction of radial coordinate is on the scattering plane and perpendicular to the velocity.

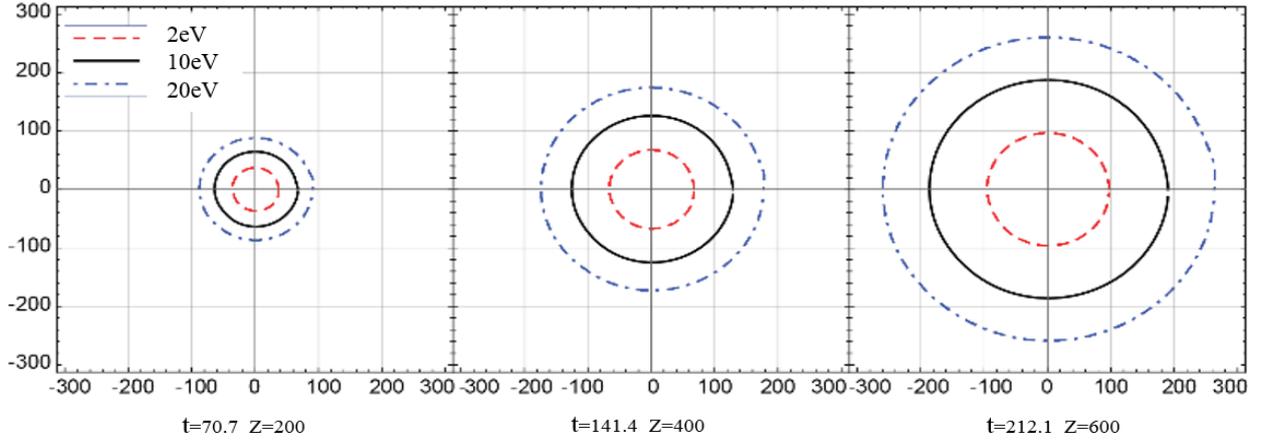

*Figure 6. Time evolution of the equal energy surface on the scattering plane. Here t is the time and Z is the z coordinate of the antiproton*

Proton collisions cause excitation, ionization, and charge transfer. Figure 3 shows the charge transferred to the proton moving along the z axis as a small peak. Here, to verify the charge transfer process, the position of the minor peak at each instant of time is shown for protons at z coordinates 272, 336, and 400. As shown in Figure 4, the position of the minor peak coincides with the position of the proton, indicating that the electron in this region moves with the proton. From the numerical solution of TDSE, the relationship between coordinates and energy is determined by formula (10). The relationship between energy and coordinates along the direction perpendicular to the z-axis on the scattering plane is shown in Figure 5. The increase in energy depending on the coordinates indicates that electrons with high energy move away from the center of the atom.

It can also be seen from Figure 6 that the equal energy surface expands as time increases. This indicates the outward movement of electrons. As shown in Figure 6, the equal energy surface is almost spherically distributed. The deviation of the energy surface from the spherical surface is not more than 3%. The relationship (10) between energy and coordinates was determined numerically from the phase of the time-dependent wave function at each coordinate point. As shown in Figure 7, our calculated FDCS is in good agreement with the relativistic [10] and quantum mechanical [7] results.

Also, our results calculated by the CWDVR method at other energies have been previously shown to be consistent with other theoretical calculations [13]. Therefore, the new expression FDCS (7) defined by the wave function is compared with the FDCS calculated by the traditional method (11).

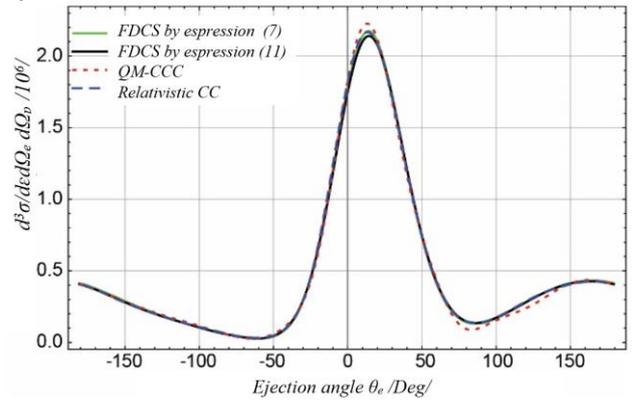

*Figure 7. When an antiproton with an energy of 200 keV is corresponding to scatering angle of 0.2 mrad, an electron FDCS ionized with an energy of 7 eV is plotted on the scattering plane. Here horizontal axis is angle $\theta_e$. Green line – FDCS by espression (7), black line – FDCS by espression (11), blue dashed line– relativistic CC [10], dots - QM-CCC [7].*

For proton-hydrogen atom collisions, using the wave function (3), the FDCS is obtained by equation (7) and compared with the results of the traditional method (11) in Fig. 8.

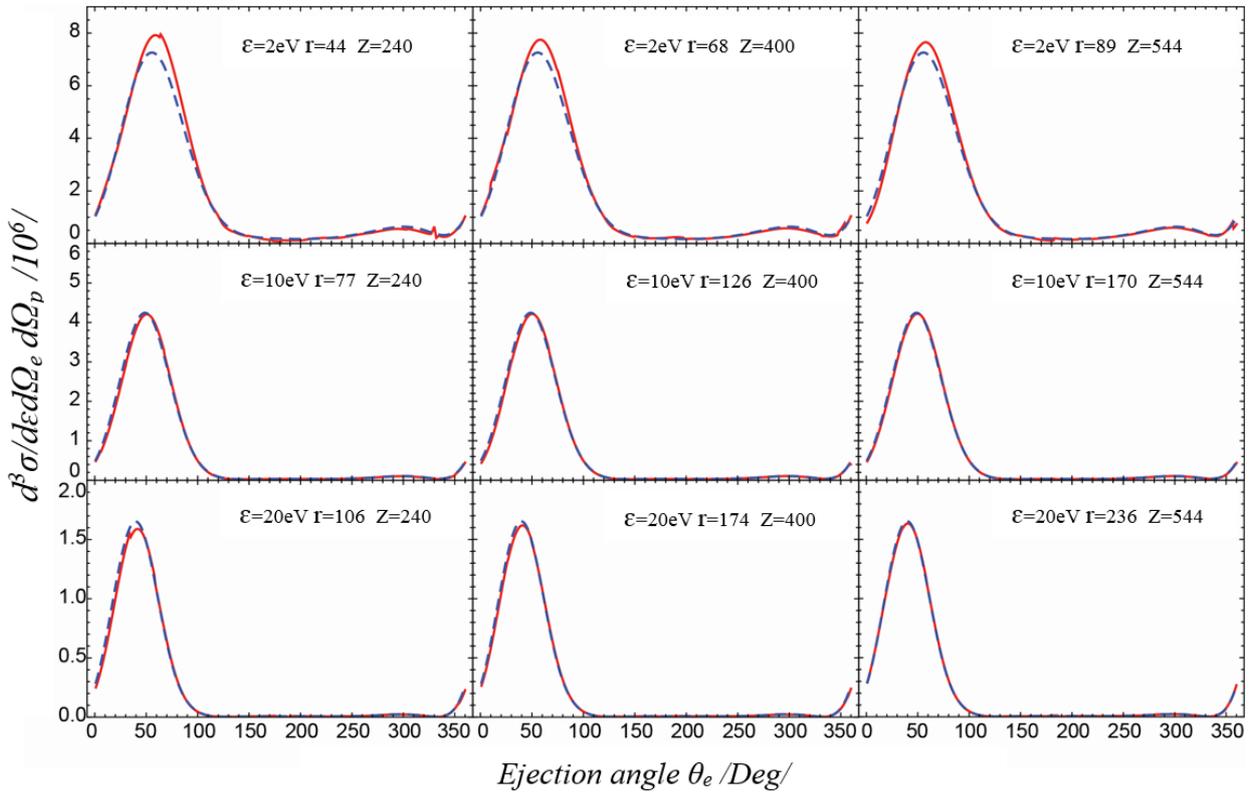

*Figure 8. The FDCS shows at three different moments of the proton z coordinates 240, 400, and 544 in the projection plane when a proton scattering angle of 0.2 mrad or a perpendicular constructor of the transfer momentum is η=0.519. Here ε is the ejection energy, z is the z-coordinate of the proton, and r is the radial coordinate of the electron. The red curve shown in the figure corresponds to the FDCS calculated by the new expression (7), and the blue dashed curve corresponds to the FDCS calculated by the traditional method (11).*

Let's evaluate the coincidence of these results. The relative difference in the maximum value of FDCS decreases with increasing time from 3.9% to 2.4% at 2 eV, from 0.9% to 0.6% at 10 eV, and from 2.8% to 0.8% at 20 eV. The position of the protons is between 240 and 544. This shows that the FDCS of the present expression (7) is in good agreement with the calculated by traditional method FDCS (11) when the ionized electron moves away from the atomic nucleus. The calculated redults of antiproton- hydrogen atomic collision is shown in Figure 9.

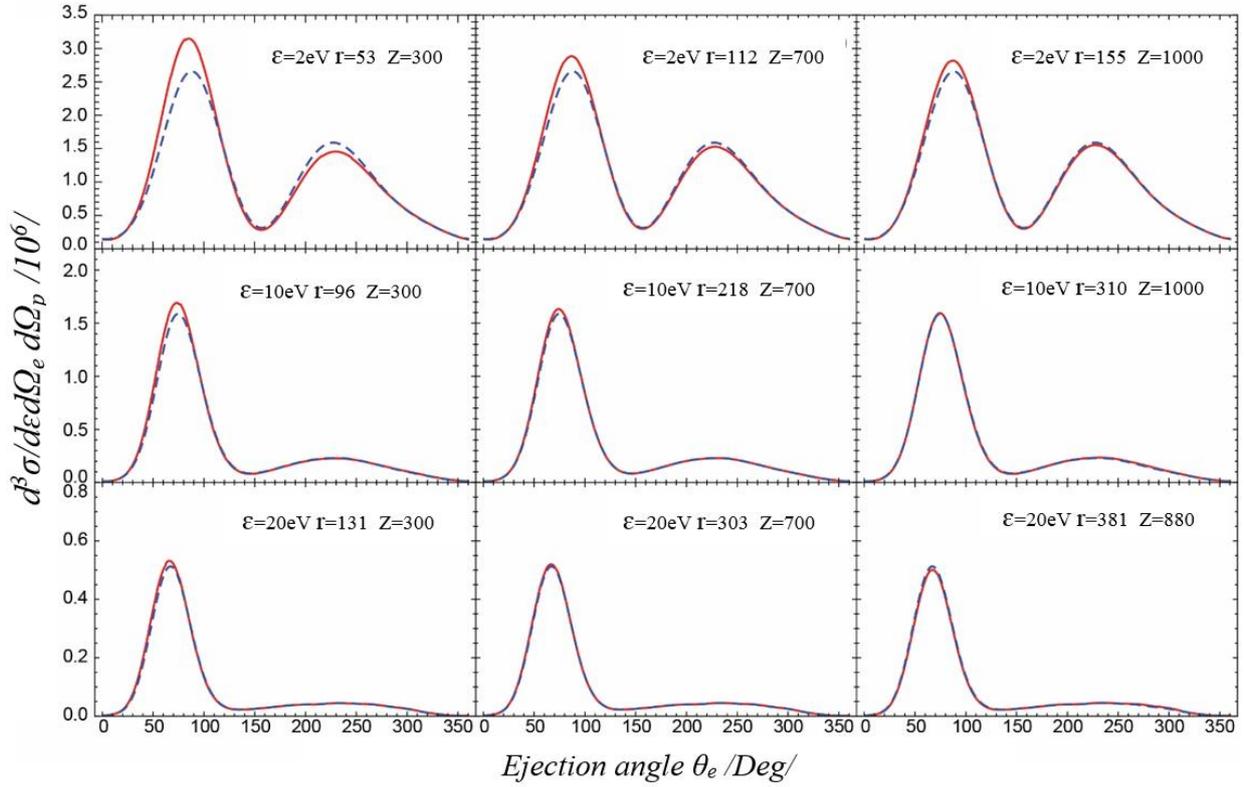

*Figure 9. The FDCS shows at three different moments of the antiproton z coordinates 240, 400, and 544 in the projection plane when a antiproton scattering angle of 0.2 mrad or a perpendicular constructor of the transfer momentum is η=0.519. Here ε is the ejection energy, z is the z-coordinate of the proton, and r is the radial coordinate of the electron. The red curve shown in the figure corresponds to the FDCS calculated by the new expression (7), and the blue dashed curve corresponds to the FDCS calculated by the traditional method (11).*

Figures 8 and 9 show that when the ejection energy is low, the FDCS results of the 2 methods take longer to converge. This can be explained by the fact that low-energy electrons move slowly away from the nucleus. The proximity of the electron to the nucleus can affect the results of FDCS. On the other hand, as the discharge energy increases, the present FDCS results converge rapidly, as shown in Figures 8 and 9 .

## IV. CONCLUSIONS

The results of FDCS (7) calculated from our newly formulated time-dependent wave function are in good agreement with the values of FDCS (11) calculated from the transition amplitude. Compared to traditional methods, this method requires calculations in a wide range of space and time, but there is no need the integral for calculating the transition amplitude. It has been shown that FDCS results recorded by a device at a macro distance can be determined using the phase and amplitude at a given point of the time-dependent wave function of a single electron. Historically, Keml's imaging theorem [20] related the probability differential in macrospace to the probability differential in momentum space, and developed a method to calculate FDCS using transition amplitudes. The possibility of determining the probability density in the macrospace by the wave function in the microspace outside the reaction zone is demonstrated by Keml's imaging theorem.

For atoms with multi-electron , FDCS calculations are difficult because the continium wave function is not analytically defined. This problem can be avoided by using the new expression FDCS (7).

## Author Contributions

All authors contributed to the study conception and design. G.Z developed an expression of fully differential cross section by the wave function. An idea of transformation of the final state electron wave function from the impact parameter to the transferred momentum representation belongs to L.Kh, Ch.A, O.A mainly worked on the developing of the code for solution of Schrödinger's equation and calculations of results. All authors read and approved the final manuscript.

**Data Availability Statement** This manuscript has no associated data or the data will not be deposited.

## V. REFERENCES

[1] S. Jones and D. H. Madison, Scaling behavior of the fully differential cross section for ionization of hydrogen atoms by the impact of fast elementary charged particles, Phys. Rev. A 65, 2002, 052727.

[2] A. B. Voitkiv and J. Ullrich, Three-body Coulomb dynamics in hydrogen ionization by protons and antiprotons at intermediate collision velocities, Phys. Rev. A 67, 2003, 062703.

[3] A. Igarashi, S. Nakazaki, and A. Ohsaki, Ionization of atomic hydrogen by antiproton impact, Phys. Rev. A, 2000, Vol 61, 062712.

[4] Xiao-Min Tong, Tsutomu Watanabe, Daiji Kato, and Shunsuke Ohtani. Ionization of atomic hydrogen by antiproton impact: A direct solution of the time-dependent Schrodinger equation .Physical review A, 2001, Volume 64, 022711.

[5] M. McGovern, D. Assafrao, J. R. Mohallem, C. T. Whelan, and H. R. J. Walters, Differential and total cross sections for antiproton-impact ionization of atomic hydrogen and helium, Phys. Rev. A, 79, 2009, 042707.

[6] M. McGovern, D. Assafrao, J. R. Mohallem, C. T. Whelan, and H. R. J. Walters, Coincidence Studies with Antiprotons, Phys. Rev. A 81, 2010, 032708.

[7] I. B. Abdurakhmanov, A. S. Kadyrov, I. Bray, and A. T. Stelbovics, Differential ionization in antiproton–hydrogen collisions within the convergent-close-coupling approach, J. Phys. B: At. Mol. Opt. Phys. 44, 2011, 165203.

[8] M. F. Ciappina, T.G. Lee, M. S. Pindzola, and J. Colgan, Nucleus-nucleus effects in differential cross sections for antiproton-impact ionization of H atoms, Phys. Rev. A 88, 2013, 042714.

[9] I. B. Abdurakhmanov, A. S. Kadyrov, and I. Bray, Wave-packet continuum-discretization approach to ion-atom collisions: Nonrearrangement scattering, Phys. Rev. A 94, 2016, 022703.

[10] A. I. Bondarev, Y. S. Kozhedub, I. I. Tupitsyn, V. M. Shabaev, G. Plunien, and Th. Stöhlker. Relativistic calculations of differential ionization cross sections: Application to antiproton-hydrogen collisions. Phys. Rev. A 95, 2017, 052709.

[11] K.M. Dunseath, J.M Launay, M Terao-Dunseath and L M ouret J. Schwartz interpolation for problems involving the Coulomb potential. Phys. B: At. Mol. Opt. Phys. 35, 2002, 3539–3556

[12] Peng. Liang-You and Starace. Anthony F, Application of Coulomb wave function discrete variable representation to atomic systems in strong laser fields. J. Chem. Phys. 125, 2006,154311

[13] G. Zorigt, L. Khenmedekh, Ch. Aldarmaa, Fully differential cross sections of proton-hydrogen and antiproton-hydrogen collisions. IJMA- 10(5), 2019, 19-23.

[14] Andrey I. Bondarev, Ilya I. Tupitsyn, Ilia A. Maltsev, Yury S. Kozhedub, and Gunter Plunien. Positron creation probabilities in low-energy heavy-ion collisions. Eur.Phys. J. D 69, 2015, 110

[15] L. D. Landay and E. M. Lifshitz, Quantum Mechanics. No relativistic Theory, 4th ed, 1989,page 53

[16] J. Azuma, N. Toshima, K. Hino, and A. Igarashi, B-spline expansion of scattering equations for ionization of atomic hydrogen by antiproton impact. Phys. Rev. A 64, 2001, 062704

[17] Emil Y Sidky and C D Lin. J. Numerical solution of the time-dependent Schrödinger equation for intermediate-energy collisions of antiprotons with hydrogen. Phys. B: At. Mol. Opt. Phys. 31, 1998, 2949–2960

[18] J. C. Wells, D. R. Schultz, P. Gavras and M. S. Pindzola. Numerical Solution of the Time-dependent Schrödinger Equation for Intermediate-Energy Collisions of Antiprotons with Hydrogen, Phys. Rev. A 54, 1996, 593

[19] B. Pons. Ability of monocentric close-coupling expansions to describe ionization in atomic collisions. Phys. Rev. A 63, 2000, 012704

[20] J S Briggs and J M Feagin. Autonomous quantum to classical transitions and the generalized imaging theorem. New J. Phys. 18 (2016) 033028